# High-pressure study on the superconducting pyrochlore oxide $Cd_2Re_2O_7$

Zenji HIROI*, Touru YAMAUCHI, Takahiro YAMADA[1], Masafumi HANAWA, Yasuo OHISHI[1], Osamu SHIMOMURA[2], Melike ABLIZ, Masato HEDO, and Yoshiya UWATOKO

*Institute for Solid State Physics, University of Tokyo, Kashiwanoha, Kashiwa, Chiba 277-8581, Japan*
[1]*Japan Synchrotron Radiation Research Institute (JASRI)/SPring-8, Sayo-gun, Hyogo 679-5198, Japan*
[2]*Japan Atomic Energy Research Institute (JAERI)/SPring-8, Sayo-gun, Hyogo 679-5148, Japan*

Superconducting and structural phase transitions in a pyrochlore oxide $Cd_2Re_2O_7$ are studied under high pressure by x-ray diffraction and electrical resistivity measurements. A rich *P-T* phase diagram is obtained, which contains at least two phases with the ideal and slightly distorted pyrochlore structures. It is found that the transition between them is suppressed with increasing pressure and finally disappears at a critical pressure $P_c$ = 3.5 GPa. Remarkable enhancements in the residual resistivity as well as the coefficient *A* of the $AT^2$ term in the resistivity are found around the critical pressure. Superconductivity is detected only for the phase with the structural distortion. It is suggested that the charge fluctuations of Re ions play a crucial role in determining the electronic properties of $Cd_2Re_2O_7$.

## §1. Introduction

Very recently superconductivity with a critical temperature $T_c \sim 1$ K was found in the pyrochlore oxide $Cd_2Re_2O_7$.[1,2,3] It is the first superconductor in the family of pyrochlore oxides which possess a unique tetrahedral network of transition metal ions. The following experimental efforts to elucidate the mechanism of the superconductivity are now being done extensively. Specific heat and Re NQR measurements have revealed that the superconducting gap is almost isotropic.[4,5] In contrast, μSR measurements have pointed out a possibility of an anisotropic superconducting order parameter.[6]

Besides superconductivity, $Cd_2Re_2O_7$ exhibits an interesting phase transition at high temperature of $T_s$ = 200 K, which may reflect the peculiar electronic structure of the compound and must be closely related to the occurrence of superconductivity at low temperature.[1,7,8] It is believed that, as temperature decreases, the crystal structure changes from the ideal cubic pyrochlore structure with space group $Fd3m$ to another slightly distorted cubic structure with space group $F\bar{4}3m$.[7] The transition is of the second order and the cell volume increases gradually with decreasing temperature below $T_s$. The structural refinements of the low-temperature (LT) phase has not yet been completed. However, a consideration on the change of space groups from $Fd3m$ to $F\bar{4}3m$ which involves the lack of inversion symmetry at the Re site leads us to assume that each Re tetrahedron can expand or shrink alternatingly without distortions, resulting in a kind of breathing mode frozen in the three-dimensional network.[7] It is considered, in other words, that all the bonds of the Re tetrahedral network are identical forming the diamond structure in the high-temperature (HT) phase, while they exhibit an alternation in the LT phase forming the zinc-blende structure.[9]

On the other hand, this structural transition is accompanied by various, dramatic changes in physical properties:[1] The resistivity is almost independent of temperature above $T_s$, while it shows a good metallic behavior below $T_s$ with an apparent kink at $T_s$. Correspondingly, magnetic susceptibility exhibits a sudden decrease below $T_s$, with the value at $T$ = 5 K reduced by about 35% compared with that at room temperature.[1] These facts mean that a crucial change in the electronic structure occurs at the transition, suggesting that the origin is purely electronic and attributable to an inherent instability in the electronic structure of the pyrochlore oxide.

Since the cell volume is larger in the LT phase, we have expected that the application of high pressure can suppress the transition and thus can stabilize the HT structure. This is the motivation of the present high-pressure (HP) study. Here we report a structural study and resistivity measurements on $Cd_2Re_2O_7$ under high pressure. An intriguing *P-T* phase diagram has been obtained, which illustrates a rich physics in the itinerant electron system on the pyrochlore lattice.

## §2. Experimental

Single crystal and polycrystalline samples were prepared from CdO, $ReO_3$ and Re in evacuated quartz ampoules at 900ºC as reported previously.[10] Powder x-ray diffraction (XRD) measurements were performed in SPring-8 on the BL10XU beam line. A polycrystalline sample was put into a diamond-anvil cell with a pressure medium of methanol and ethanol in a ratio of 4:1 and cooled down to *T* = 14 K. Debye-Scherrer ring patterns were collected during isothermal compression at 14 K up to *P* = 13 GPa, using a monochromated synchrotron radiation beam with a wavelength of 0.04964 nm.

Electrical resistivity measurements under HP were carried out using a standard four-probe method in a cubic-anvil type apparatus which enabled us to apply a hydrostatic pressure up to 8 GPa. A single crystal with dimensions of 0.5 x 0.16 x 0.06 mm$^3$ was immersed into a flourinate which was used as a pressure medium in a pressure cell. Each measurement was carried out at certain pressure on cooling and then on heating between 2.3 K and 300 K, with keeping pressure load to the cubic-anvil cell always constant. The pressure was varied at $T$ = 300 K between the measurements.

## §3. Results

### 3.1 X-ray Diffraction Measurements

A powder XRD pattern taken at $P$ = 2.1 GPa and $T$ = 14 K could be indexed with a cubic unit cell, and the extinctions were compatible with space group $F\bar{4}3m$ reported for the LT phase at ambient pressure (AP): Four weak reflections 002, 024, 006, and 046 which are forbidden in the HT phase were detected, as observed in a previous single-crystal XRD measurement[7] but not in ordinary powder XRD experiments using a rotating anode diffractometer. Figure 1 shows the evolution of the 002 and 046 reflections as a function of pressure: The 002 reflection obviously disappears above a critical pressure $P_c$ between 2.8 and 5.3 GPa. A weak trace of intensity above $P_c$ may come from a background. On the other hand, the 046 reflection also becomes indiscernible around a similar pressure, although considerable broadening of the nearby fundamental reflection 515 makes it ambiguous. This peak broadening develops at higher pressure and may be due to inhomogeneous distribution in pressure inside the pressure cell. The observed disappearance of the 00$l$ and 0$kl$ type reflections means that a pressure-induced structural transition from $F\bar{4}3m$ to $Fd\bar{3}m$, the latter being the same as in the HT phase at AP. However, we do not exclude the possibility for other space groups, because our experiments are not so conclusive for the lack of the 0$kl$ type reflection because of the peak broadening. If it survives at HP, a possible space group would be $F4_132$ which also belongs to the subgroup of $Fd\bar{3}m$.

We have refined the cell parameter by fitting the whole pattern of $2\theta$ = 4-38° ($d$ = 0.6 - 0.075 nm) using the Rietveld method. The pressure dependence of the cell volume $V = a^3$ is shown in Fig. 1(b). Anomalies to be seen at the transition are not clear, suggesting that the pressure-induced transition is of the second order. As reported previously, an isostructural osmium pyrochlore oxide $Cd_2Os_2O_7$ keeps the $Fd\bar{3}m$ structure down to 10 K in spite of a distinct metal-semiconductor transition at $T$ = 230 K.[7,11] The temperature dependence of the cell parameter is continuous without any anomalies at the transition. Thus, it must exhibit an intrinsic thermal expansion of the ideal pyrochlore structure. Using the data from $Cd_2Os_2O_7$, we estimated the cell parameter of the ideal pyrochlore structure for $Cd_2Re_2O_7$ at LT below $T_s$ which would be attained if the transition is suppressed in some way. It is $a_0$ = 1.0216 nm ($V_0$ = 1.0662 nm$^3$) at $T$ = 14 K. Taking this value and the data well above $P_c$ ($P \geq 6$ GPa), we obtained the pressure dependence of the cell volume for $Cd_2Re_2O_7$ in the ideal pyrochlore structure by fitting the data to the form $\Delta V/V_0 = -\alpha P + \beta P^2$, where $\alpha$ is the compressibility. The result is shown in Fig. 1(b) with $\alpha$ = 4.3 x 10$^{-3}$ (GPa)$^{-1}$ and $\beta$ = 1.2 x 10$^{-4}$ (GPa)$^{-2}$. It is to be noted that the data points lie slightly above the curve below $P_c \sim 4$ GPa, and the change at $P_c$ may be continuous.

### 3.2 Resistivity Measurements

Resistivity $\rho$ measured at various pressures is shown in Fig. 2. The crystal used shows at AP $T_c$ = 0.98 K, and $\rho$ = 400 $\mu\Omega$cm and 26 $\mu\Omega$cm at $T$ = 300 K and 5 K, respectively. Thus the residual resistivity ratio (RRR) is about 15, which is smaller than the largest RRR so far we obtained (about 30).[1] However, this may not cause a serious problem, because the HT phase transition as well as the superconducting transition are rather insensitive to the RRR value.[4] As reported previously,[1] the resistivity shows a distinct kink at $T_s$, and thus the $T_s$ is determined as a temperature where the derivative of the resistivity curve exhibits a peak. Obviously seen in Fig. 2 is the shift of the kink to lower temperature with increasing pressure; $T_s$ = 200 K, 143 K, and 80 K for $P$ = AP, 1.5 GPa, and 3.0 GPa, respectively. This implies that the HT phase is stabilized in fact under HP. The transition does not tend to smear out under HP. The resistivity drop is even more distinct at $P$ = 1.5 GPa than at AP. This fact indicates a homogeneous pressure distribution in the present experiments. Generally, more homogeneous pressure is attained in a multi-anvil type HP apparatus as used in the present resistivity measurements than in a diamond-anvil cell as used in the present XRD experiments.

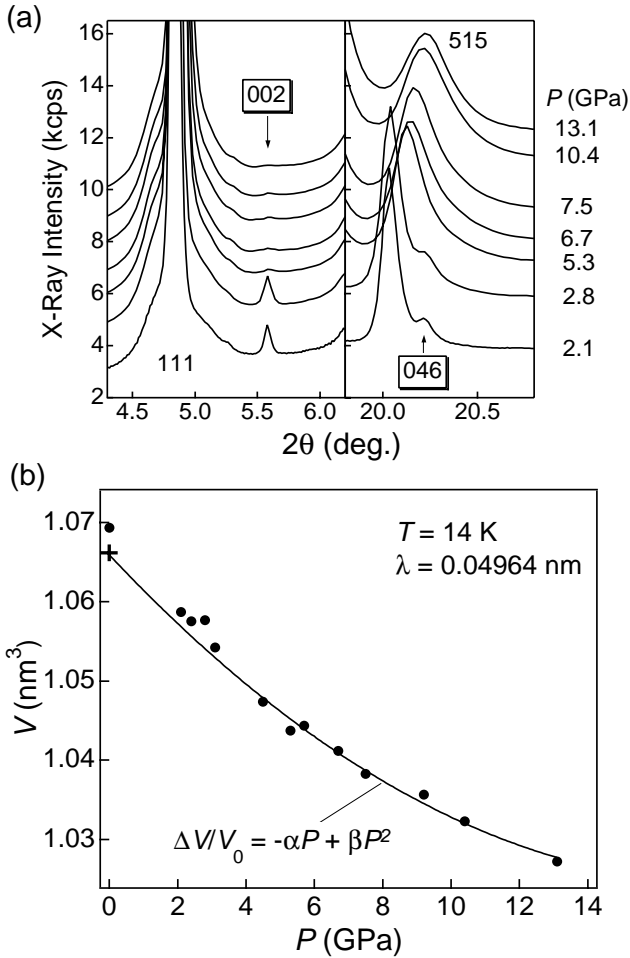

Fig. 1. (a) X-ray diffraction patterns of $Cd_2Re_2O_7$ taken at $T$ = 14 K and various pressures, showing the 111 and 002 reflections (left panel) and the 515 and 046 reflections (right panel). Both the 002 and 046 reflections are not allowed to appear for space group $Fd\bar{3}m$ but for $F\bar{4}3m$. (b) Pressure dependence of the cell volume. The cross represents $V_0$, and the solid line does a fit to the equation $\Delta V/V_0 = \alpha P + \beta P^2$. See text for detail.

It is to be noted that the $\rho$ above $T_s$ always trace the same curve for $P \leq 3.0$ GPa, and it is the curve for $P = 3.5$ GPa. In other words the $P$ dependence of $\rho$ at $T > T_s$ is nearly zero, while it is significantly large with a negative sign for $T < T_s$. This remarkable pressure independence for the HT phase suggests an unusual electronic structure underlying. An obvious resistive transition is not seen for $P = 3.5$ GPa, but the $\rho$ decreases gradually below $T \sim 50$ K. When pressure is further increased, as shown in Fig. 2(b), the $\rho$ curve shifts downward in a wide temperature range. Therefore, we have concluded that the critical pressure $P_c$ is close to 3.5 GPa.

At $P > P_c$ we have noticed another small anomaly at $T^*$ which is shifted to higher temperature with increasing pressure: $T^* = $ 145 K, 175 K, 195 K, and 230 K at $P = 4.0$, 5.0, 6.0, and 8.0 GPa, respectively. Although this anomaly is less distinct compared with that at $T_s$, it is well recognized as a point where the pressure dependence of $\rho$ changes. For example, at $P = 4.0$ GPa, the $\rho$ above $T^*$ is nearly the same as that of $P = 3.5$ GPa, but apparently the two curves deviate from each other below $T^*$. Possibly there is another phase transition or a crossover at $T^*$ and $P > P_c$. One more notable feature in Fig. 2 is anomalous enhancement in the residual resistivity $\rho_0$ toward $P_c$. The $\rho_0$ value at $P = 3.5$ GPa is about one order larger than that at AP. Its pressure dependence is summarized in the inset to Fig. 3, and will be discussed later.

Concerning the temperature dependence of resistivity at low temperature above $T_c$, figure 3 plots the $\rho-\rho_0$ against $T^2$ below $T = 20$ K. As mentioned previously,[1]) the temperature dependence of $\rho$ at AP is proportional to $T^3$, not $T^2$, below $T = 25$ K, suggesting that electron correlations are less dominant compared with other scattering mechanisms such as a phonon. A $T^2$ behavior is only seen below $T \sim 10$ K. When pressure is applied, however, the initial slope of the $\rho$-$T^2$ curve gradually increases, and the temperature window where the $T^2$ behavior is attained becomes widened, as shown in Fig. 3. Around $P_c$ these variations tend to be saturated and then return with further increasing pressure. The $\rho-\rho_0$ plotted against $T^3$ showed a convex curve at $P = P_c$. The variation of the coefficient $A$ of the $T^2$ term is plotted in the inset to Fig. 3. Surprisingly observed is a pronounced peak at $P = P_c$ and a scaling behavior between the $A$ and $\rho_0$. In an ordinary Fermi liquid picture, the $A$ is proportional to $(m^*)^2$, where the $m^*$ is an effective mass of carriers. Since the $A$ becomes one order larger at $P_c$ than at AP, the $m^*$ of phase I must be much heavier than in phase II, approximately three times larger. This is qualitatively in good agreement with the experimental fact that the density of state (DOS) estimated from magnetic susceptibility measurements is about 1.5 times larger in the HT phase than in the LT phase.

At ambient pressure $Cd_2Re_2O_7$ exhibits bulk superconductivity at $T_c = 1.0$ K. It is interesting to know how pressure affects the $T_c$ and what kind of ground states would appear when the superconductivity is suppressed. Figure 4 shows resistivity data taken at low temperature down to 2.3 K. We found a distinct zero-resistive transition at $T = 2.7 \sim 3.0$ K only at $P = 1.5$, 2.0, and 2.5 GPa. The pressure dependence of $T_c$ is plotted in Fig. 5. In contrast, at higher pressures, there is a tiny anomaly in $\rho$ without a zero-resistive state. It seems that the superconducting state is confined in the LT phase region. However, since our data is limited above 2.3 K, it is not known whether the $T_c$ increases continuously from 1.0 K at AP to 3.0 K at $P \sim 2$ GPa. Moreover, it should be examined how the $T_c$ varies near the phase boundary at $P = 3.5$ GPa. We plan to do another HP experiments at lower temperature under HP.

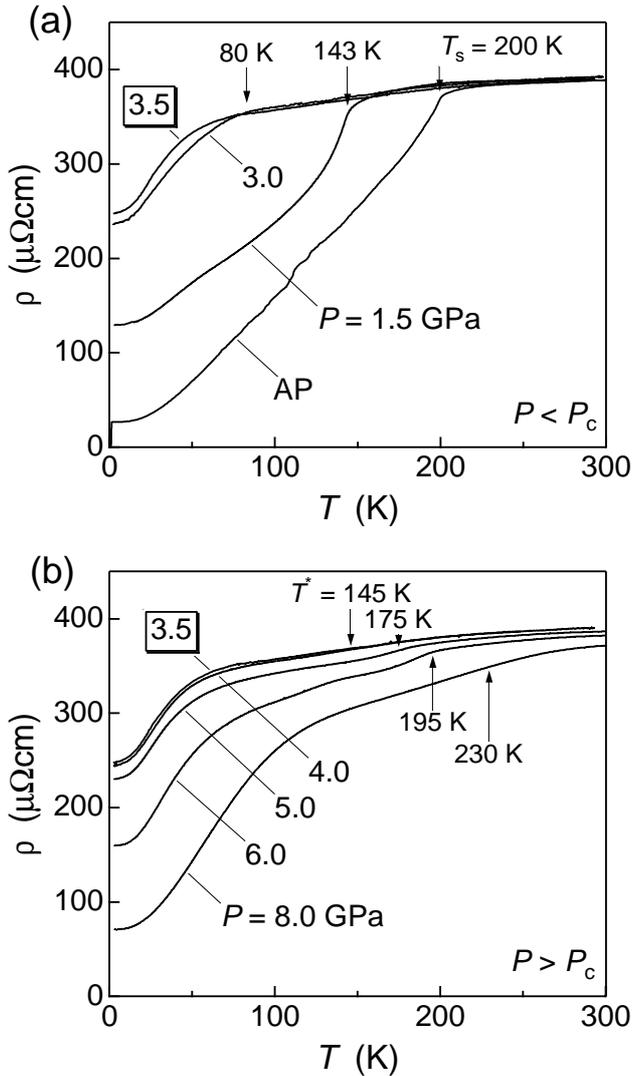

Fig. 2. Temperature dependence of resistivity $\rho$ measured under pressures below 3.5 GPa (a) and above 3.5 GPa up to 8 GPa (b). Structural transition temperature $T_s$ and another characteristic temperature $T^*$ are marked by arrows in (a) and (b), respectively.

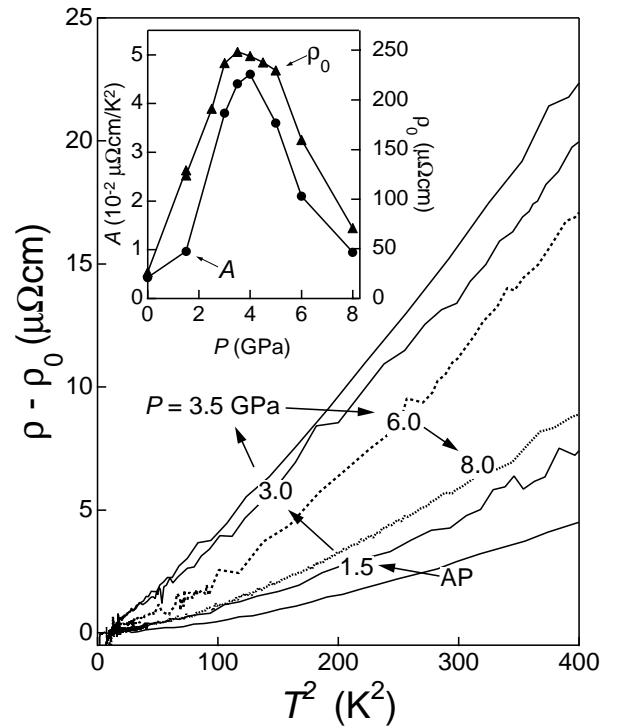

Fig. 3. Resistivity ($\rho - \rho_0$) at low temperature below 20 K is plotted against $T^2$. The $T^2$ term in the temperature dependence becomes dominant near $P = 3.5$ GPa. The pressure dependence of its coefficient $A$ and $\rho_0$ are shown in the inset.

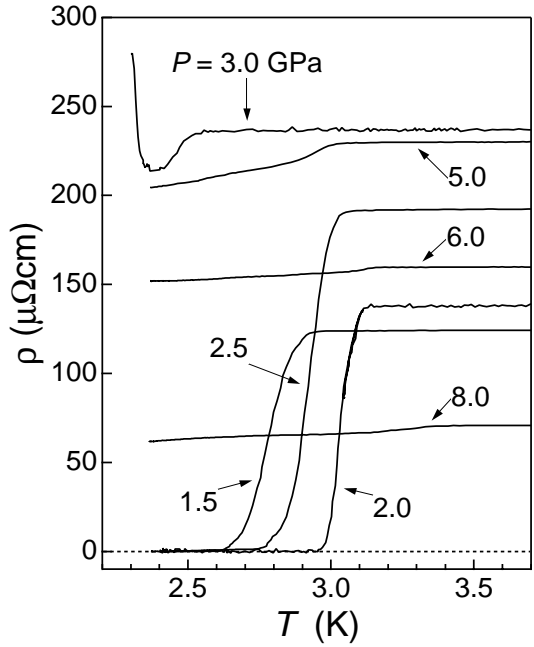

Fig. 4. Resistivity at low temperature showing a superconducting transition at $T_c = 2.7$-$3.0$ K only for $P = 1.5$, $2.0$, and $2.5$ GPa.

*3.3 P-T Phase Diagram*

Figure 5 summarizes the result of the present HP study on $Cd_2Re_2O_7$ in the $P$-$T$ phase diagram. There are at least two phases at AP: Phase I is a cubic $Fd3m$ phase with the ideal pyrochlore structure, and phase II has a slightly distorted pyrochlore structure which probably belongs to the cubic $F\bar{4}3m$ space group. The transition between them is of the second order. Applying pressure dramatically reduces the transition temperature $T_s$ with a large initial slope of $dT_s/dP = -33$ K/GPa. The transition steeply vanishes at $P_c = 3.5$ GPa. The HP XRD experiments performed at $T = 14$ K consistently indicate that a structural phase transition occurs at a similar critical pressure. At $P > P_c$ another specific temperature $T^*$ appears. If this corresponds to a real phase transition, there should be the third phase named I'. In this case it is plausible that the space group of phase I' is different from $Fd3m$ and $F\bar{4}3m$, having intermediate symmetry, such as $F4_132$. It is to be noted that the residual resistivity $\rho_0$ exhibits a pronounced peak with the maximum at $P_c$, as shown in the inset to Fig. 3.

Electronic structures must be quite different among these phases. Phase I is the most unconventional phase where the resistivity is almost independent of both temperature and pressure. The $\rho$ curve observed at $P = P_c$ should indicate the intrinsic temperature dependence of resistivity for phase I. The anomalously large residual resistivity is not apparently due to simple impurity scattering but may reflect a unique scattering process involved in this phase with undistorted pyrochlore structure. In contrast phases II and I' seem to have more ordinary electronic structures showing normal temperature and pressure dependences of resistivity.

A superconducting ground state may exist only for phase II. It seems that the distortion of the pyrochlore lattice and thus the breakdown of the high degeneracy are necessary for the occurrence of superconductivity. In this sense frustration may be harmful to superconductivity. This can be one of the reasons why superconductivity has been found rarely in the family of pyrochlore oxides.

§4. Discussion

According to the recent electronic structure calculations, $Cd_2Re_2O_7$ (phase I) is a semi-metal with low carrier density.[12,13] There are hole bands at the $K$ point and electron bands at the $\Gamma$ point in the Brillouin zone. Because of high degeneracy at the $K$ point, it is expected that the hole bands have larger contributions to the DOS than the electron bands.[13] The carrier density calculated is $2.5 \times 10^{-3}$ per Re atom for each band.[13] The low carrier density and the less dispersive bands lead $Cd_2Re_2O_7$ to a strongly correlated electron system, which is not due to ordinary on-site Coulomb repulsion but due to $k$ dependent correlations near the Fermi level.[12,13] Unfortunately, similar calculations for phase II are not available at present. However, based on the considerations on the symmetry change from $Fd3m$ to $F\bar{4}3m$, Harima predicted that the lack of inversion symmetry in $F\bar{4}3m$ lifts spin degeneracy, and that only the hole bands at the $K$ point splits into two so that one hole sheet disappears.[13] This can explain the large reduction in magnetic susceptibility below $T_s$ and the large mass enhancement near $P_c$ observed in the present study.

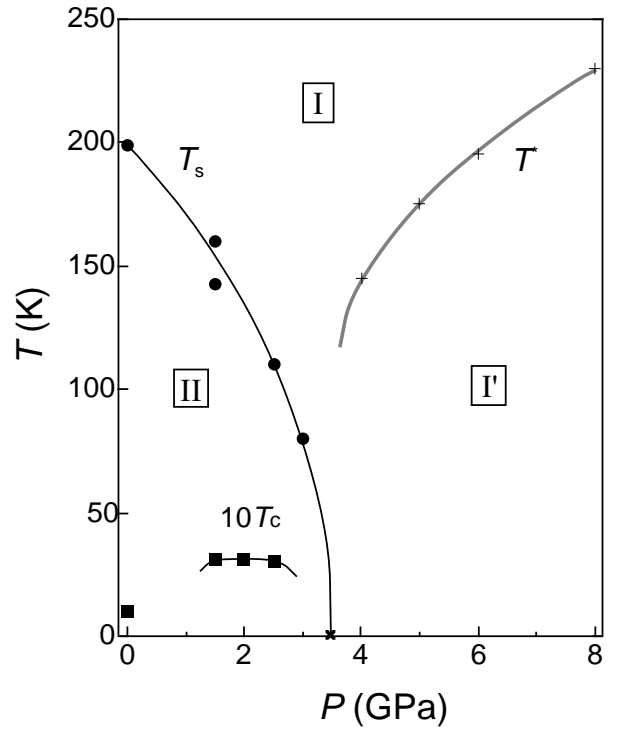

Fig. 5. $P$-$T$ phase diagram of $Cd_2Re_2O_7$. Phase I represents a high-temperature form with the ideal pyrochlore structure (space group $Fd3m$) and phase II does a low-temperature form with slightly distorted pyrochlore structure (possibly space group $F\bar{4}3m$). The transition between them is suppressed with increasing pressure and finally disappears at a critical pressure $P_c = 3.5$ GPa. Another phase transition or crossover at $T^*$ exists for $P > P_c$, below which phase I' may exist. A superconducting transition above $T = 2.3$ K is observed only within the phase I region around $P = 2$ GPa. The $T_c$ multiplied by 10 is plotted together with the $T_c = 1$ K at ambient pressure.

The origin of the observed large enhancement of $\rho_0$ near the phase boundary at $T = 0$ is very interesting and would provide a key to understand the physics involved on this pyrochlore compound. A similar anomaly in $\rho_0$ have been observed in heavy fermion compounds like $CeCu_2Ge_2$ and $CeCu_2Si_2$ where the $A$ rapidly decreases and the $\rho_0$ exhibits a sharp peak at $P \sim 16$ GPa after the antiferromagnetic phase is suppressed under HP.[14] Superconductivity appears at $P > 10$ GPa in $CeCu_2Ge_2$ and its $T_c$ shows a maximum at $P \sim 16$ GPa. The relation between the $\rho_0$ and $T_c$ has been discussed in terms of enhanced valence fluctuations of Ce.[15,16] It is considered in the present compound that the enhanced $\rho_0$ near the phase boundary indicates inherent charge fluctuations of Re. It is known that $Re^{5+}$ is not stable and is rarely seen in oxides, compared with $Re^{4+}$ or $Re^{6+}$.[17] Structural deformations induced below $T_s$ strongly suggest a sort of "charge ordering", resulting in the disproportionation of the Re tetrahedra. This charge ordering should be clearly distinguished from those seen in vanadates or manganites, because it is not coupled with the metal-to-insulator transition but with better conductivity. It is plausible that, when pressure suppresses the charge ordering to $T = 0$, relevant charge fluctuations are enhanced, giving rise to a large enhancement in $\rho_0$. The role of charge fluctuations on superconductivity, particularly superconductivity with increased $T_c$ around $P \sim 2$ GPa, might be intriguing to be explored.

§4. Concluding Remarks

In conclusion we have measured the x-ray diffraction and electrical resistivity of a superconducting pyrochlore oxide $Cd_2Re_2O_7$ under high pressure. Remarkably intimate relation between the crystal and electronic structures is revealed, which illustrates an interesting physics for itinerant electrons on the pyrochlore lattice.

It is important to determine the true crystal structure for the LT phase II. Although our previous powder and single-crystal XRD experiments indicated cubic symmetry within the experimental resolution, very recent $^{185/187}$Re NQR and $^{111}$Cd NMR experiments suggested the lack of three-fold symmetry at the metal positions, implying that the crystal system is not cubic.[5] Vyaselev *et al.* have speculated that the charge distribution associated with the Re 5$d$ orbitals is slightly distorted possibly due to orbital order, and induced lattice distortion, which is too small to be detected in ordinary XRD experiments.[5] Further structural analyses using a high-resolution synchrotron radiation and a neutron source are now in progress.

Acknowledgments

We would like to thank H. Harima and K. Miyake for valuable comments and discussions, and M. Takigawa, M. Sigrist, K. Ueda for helpful discussions during the course of this study. This research was supported by a Grant-in-Aid for Scientific Research on Priority Areas (A) and a Grant-in-Aid for Creative Scientific Research given by The Ministry of Education, Culture, Sports, Science and Technology, Japan. The synchrotoron radiation experiments were performed at the SPring-8 with the approval of the Japan Synchrotoron Radiation Research Institute (JASRI) (Proposal No. 2001A0262-ND-np).